\documentclass[pra,floatfix,showpacs,amsmath,twocolumn]{revtex4}

\usepackage{amssymb}
\usepackage[mathscr]{eucal}
\usepackage[dvips]{graphicx}

\newcommand{\tr}{\textrm{tr}}

\newcommand{\ket}[1]{| \, #1 \, \rangle}
\newcommand{\bra}[1]{\langle \, #1 \, |}

\newcommand{\scal}[2]{\bra{#1} \, #2 \, \rangle}
\newcommand{\expect}[1]{\langle #1 \rangle}

\newcommand{\adj}[1]{#1^{\dagger}}

\newcommand{\sx}[1]{\sigma_x^{(#1)}}
\newcommand{\sy}[1]{\sigma_y^{(#1)}}
\newcommand{\sz}[1]{\sigma_z^{(#1)}}

\begin{document}

\title{Variational study of hard-core bosons in a 2--D optical lattice using Projected Entangled Pair States (PEPS)}

\author{V. Murg$^1$, F. Verstraete$^2$, J. I. Cirac$^1$}
\affiliation{$^1$Max-Planck-Institut f\"ur Quantenoptik,
Hans-Kopfermann-Str. 1, Garching, D-85748, Germany\\
$^2$Fakult\"at f\"ur Physik, Universit\"at Wien, Boltzmanngasse 3,
A-1090 Wien}
\date{\today}

\begin{abstract}
We have studied the system of hard--core bosons on a 2--D optical
lattice using a variational algorithm based on projected
entangled-pair states (PEPS). We have investigated the ground
state properties of the system as well as the responses of the
system to sudden changes in the parameters. We have compared our
results to mean field results based on a Gutzwiller ansatz.
\end{abstract}

\maketitle


\section{Introduction}

Systems of interacting bosons in optical lattices have attracted a
lot of interest in the last few years due to recent experimental
achievements~\cite{greiner02,stoeferle04,paredes04,tolra04,greiner01,moritz03}.
In these systems, atoms are trapped by the combination of a
periodic and a harmonic potential created by counterpropagating
laser--beams. These systems of trapped atoms resemble a crystal in
the sense that atoms are localized at periodic locations. The
theoretical study of the static and dynamic properties of these
systems is quite challenging. An exact solution can only be
obtained in one dimension in the so--called Tonks--Girardeau limit
via fermionization~\cite{girardeau60,paredes04}. Outside this
limit and in higher dimensions, approximate methods have to be
used. The most common of those, the mean field approximation based
on a Gutzwiller ansatz~\cite{jaksch02}, is unfortunately known to
give imprecise predictions for correlations. The most powerful
numerical methods such as the Density Matrix Renormalization Group
(DMRG)~\cite{white92} and Quantum Monte Carlo (QMC) are also of
limited use: DMRG is mainly restricted to 1--D systems and QMC
suffers from the sign--problem as time--evolutions are
investigated.

In this paper, we apply the algorithm introduced
in~\cite{verstraetecirac04}. This algorithm is a variational
method within a class of states termed Projected Entangled--Pair
States (PEPS). It has been proven to work well for the Heisenberg
antiferromagnet~\cite{verstraetecirac04} and the frustrated
Shastry-Sutherland model~\cite{isacsson06}. The system we focus on
is the system of hard--core bosons in a 2--D optical lattice. This
model captures the essential physics of bosons in optical
lattices~\cite{aizenman04}. We use the algorithm to determine the
ground state properties of the system and to study the responses
of the system to sudden changes in the parameters. We compare our
results to mean field results based on the Gutzwiller ansatz. We
find that the PEPS and the Gutzwiller ansatz deviate clearly in
the prediction of the ground state momentum distribution and the
time--evolution of the condensed fraction of the particles.

The paper is outlined as follows: In section~\ref{sec:algorithm}
we give a brief overview of the algorithm. We specify the model we
want to investigate in section~\ref{sec:model} and explain the way
the algorithm is applied to this model in
section~\ref{sec:application}. The results of the numerical
calculations are presented in sections~\ref{sec:groundstate}
and~\ref{sec:dynamics}. We conclude with the discussion of the the
performance and the stability of the algorithm in
section~\ref{sec:performance}.


\section{The Algorithm} \label{sec:algorithm}

Although the algorithm has already been outlined
in~\cite{verstraetecirac04}, we reiterate it here in detail -
specialized for our purpose. The algorithm is a variational method
with respect to the class of PEPS. These states have been found to
be adequate for representing the ground state of numerous
many--body systems. Also, these states are favorable to
variational calculations because they possess an internal
refinement parameter, the virtual dimension~$D$, that allows to
control the precision of the calculation. While $D=1$ specializes
the PEPS to a product state, the choice $D=d^M$ (with~$M$ being
the total number of lattice sites and~$d$ the dimension of one
subsystem) enlarges the space of PEPS to the complete
Hilbert--space of the system. The purpose of the algorithm is - in
our case - to simulate the time--evolution of a system within the
subset of PEPS with a fixed~$D$. This means that after each
time--evolution step the state of the system is approximated by
the "nearest" PEPS with virtual dimension~$D$. The key element of
the algorithm is thus the optimization of the parameters of a PEPS
such that its distance to a given state is minimized.

\begin{figure}[t]
    \begin{center}
        \includegraphics{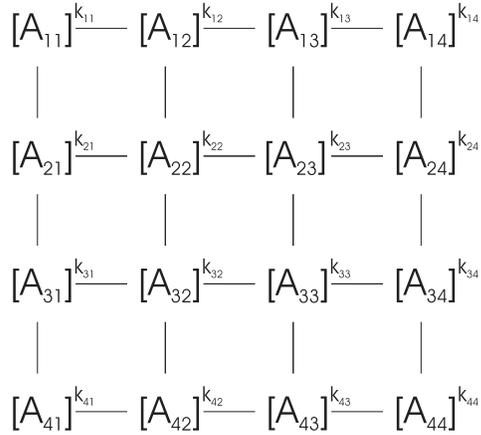}
    \end{center}
    \caption{
        Structure of the coefficient related to the state
        $\ket{k_{11},..,k_{44}}$ in the PEPS $\ket{\Psi_A}$.
    The bonds represent the indices of the tensors $[A_i]^k$
    that are contracted.
        }
    \label{fig:structpeps}
\end{figure}

The manner in which the optimization is performed is closely
related to the structure of PEPS. A PEPS is a state with
coefficients that are contractions of tensors according to a
certain scheme. Thereby, each tensor is associated with a physical
subsystem. The contraction--scheme mimics the underlying lattice
structure. Each tensor possesses one physical index with dimension
equal to the physical dimension~$d$ of a subsystem and a certain
number of virtual indices with dimension~$D$. The number of
virtual indices is equal to the number of bonds that emanate from
the lattice--site the tensor is associated with. For example, in a
rectangular lattice the number of virtual indices is~$4$ (except
at the borders) - related to the left, right, upper and lower bond
respectively. The tensor associated with site~$i$ is
\begin{displaymath}
\big[ A_{i} \big]^k_{lrud}
\end{displaymath}
with physical index~$k$ and virtual indices~$l$, $r$, $u$ and~$d$.
The coefficients of the PEPS are then formed by joining the
tensors in such a way that all indices related to same bonds are
contracted. This is illustrated in fig.~\ref{fig:structpeps} for
the special case of a $4 \times 4$ square lattice. Assuming this
contraction of tensors is performed by the function~$\mathcal{F}
(\cdot)$, the resulting PEPS can be written as
\begin{displaymath}
\ket{\Psi_A} = \sum_{k_1,...,k_M=1}^d \mathcal{F} \big( \big[ A_1
\big]^{k_1},...,\big[ A_M \big]^{k_M} \big) \ket{k_1,...,k_M}.
\end{displaymath}

The aim of the algorithm is to optimize the tensors $A_i$ such
that the distance between the PEPS~$\ket{\Psi_A}$ and a given
state tends to a minimum. We assume that the given state is a PEPS
$\ket{\Psi_B}$ with virtual dimension~$D_B$ and tensors $B_i$.
This is no loss of generality since every state can be written as
a PEPS. The function to be minimized is then
\begin{displaymath}
K \big( A_1,...,A_M \big) = \big\| \ket{\Psi_A} - \ket{\Psi_B}
\big\|^2.
\end{displaymath}
This function is non--convex with respect to all parameters
$\{A_1,...,A_M\}$. However, due to the special structure of PEPS,
it is quadratic in the parameters~$A_i$ associated with \emph{one}
lattice--site~$i$. Because of this, the optimal parameters~$A_i$
can simply be found by solving a system of linear equations. The
concept of the algorithm is to do this one--site optimization
site-by-site until convergence is reached.

The challenge that remains is to calculate the coefficient matrix
and the inhomogenity of the linear equations--system. In
principle, this is done by contracting all indices in the
expressions for the scalar--products $\scal{\Psi_A}{\Psi_A}$ and
$\scal{\Psi_A}{\Psi_B}$ except those connecting to ~$A_i$. By
interpreting the tensor $A_i$ as a $d D^4$-dimensional vector
$\boldsymbol{A}_i$, these scalar--products can be written as
\begin{eqnarray}
\label{eqn:scalaa}
\scal{\Psi_A}{\Psi_A} & = & \adj{\boldsymbol{A}_i} \mathcal{N}_i
\boldsymbol{A}_i \\
\label{eqn:scalab}
\scal{\Psi_A}{\Psi_B} & = & \adj{\boldsymbol{A}_i} \mathcal{W}_i.
\end{eqnarray}
Since
\begin{displaymath}
K = \scal{\Psi_B}{\Psi_B} + \scal{\Psi_A}{\Psi_A} - 2 Re
\scal{\Psi_A}{\Psi_B},
\end{displaymath}
the minimum is attained as
\begin{displaymath}
\mathcal{N}_i \boldsymbol{A}_i = \mathcal{W}_i.
\end{displaymath}
The obstacle, however, is that the numerical calculation of the
coefficient matrix $\mathcal{N}_i$ and the inhomogenity
$\mathcal{W}_i$ requires a number of operations that scales
exponentially with the number of subsystems~$M$. This will make
the algorithm non--efficient as the system grows larger. Because
of this, an approximate method has to be used to calculate
$\mathcal{N}_i$ and $\mathcal{W}_i$.

\begin{figure}[t]
    \begin{center}
        \includegraphics{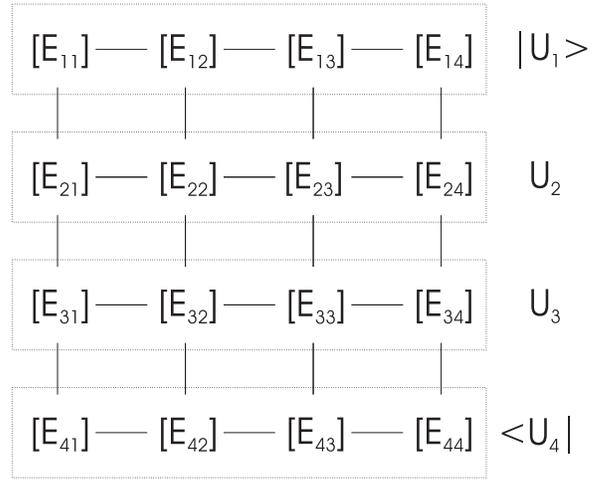}
    \end{center}
    \caption{
        Structure of the contractions in $\scal{\Psi_A}{\Psi_A}$.
    In this scheme, the first and last rows can be interpreted as
    MPS $\ket{U_1}$ and $\bra{U_4}$ and the rows in between
    as MPO $U_2$ and $U_3$. The contraction of all tensors is then
    equal to $\bra{U_4} U_3 U_2 \ket{U_1}$.
        }
    \label{fig:structe}
\end{figure}

The approximate method suggested in~\cite{verstraetecirac04} is
based on matrix product states (MPS) and matrix product operators
(MPO). To see how MPS and MPO implicitly appear in the problem of
calculating $\mathcal{N}_i$ and $\mathcal{W}_i$, we take a closer
look at the structure of the contractions in the scalar--products
$\scal{\Psi_A}{\Psi_A}$ and $\scal{\Psi_A}{\Psi_B}$. Thereby, we
focus on a $L \times L$ square lattice in the following. We start
with the study of $\scal{\Psi_A}{\Psi_A}$. For this, we single out
a specific site~$j$ and define the $D^2 \times D^2 \times D^2
\times D^2$--tensor
\begin{displaymath}
\big[ E_j \big]_{(ll')(rr')}^{(uu')(dd')} = \sum_{k=1}^d \big[
A_j^* \big]_{lrud}^k \big[ A_j \big]_{l'r'u'd'}^k.
\end{displaymath}
In this definition, the symbols $(ll')$, $(rr')$, $(uu')$ and
$(dd')$ indicate composite indices. We may interpret the~$4$
indices of this tensor as being related to the~$4$ bonds emanating
from site~$j$ in the lattice. Then, $\scal{\Psi_A}{\Psi_A}$ is
formed by joining all tensors $E_j$ in such a way that all indices
related to same bonds are contracted -- as in the case of the
coefficients of PEPS. These contractions have a rectangular
structure, as depicted in fig.~\ref{fig:structe}. In terms of the
function $\mathcal{F}(\cdot)$, the scalar--product reads
\begin{displaymath}
\scal{\Psi_A}{\Psi_A} = \mathcal{F} \big( E_1,...,E_M \big).
\end{displaymath}
The main idea of the approximate algorithm is to interpret the
first and the last row in this contraction--structure as MPS and
the rows in between as MPO. The horizontal indices thereby form
the virtual indices and the vertical indices are the physical
indices. Thus, the MPS and MPO have both virtual dimension and
physical dimension equal to~$D^2$. Explicitly written, the MPS
read
\begin{eqnarray*}
\ket{U_1} & = & \sum_{\tilde{d}_1,...,\tilde{d}_L=1}^{D^2} \tr
\Big( \big[E_{11}\big]^{1 \tilde{d}_1} \cdots \big[E_{1L}\big]^{1
\tilde{d}_L} \Big)
\ket{\tilde{d}_1,...,\tilde{d}_L}\\
\bra{U_L} & = & \sum_{\tilde{u}_1,...,\tilde{u}_L=1}^{D^2} \tr
\Big( \big[E_{L1}\big]^{\tilde{u}_1 1} \cdots
\big[E_{LL}\big]^{\tilde{u}_L 1} \Big)
\bra{\tilde{u}_1,...,\tilde{u}_L}
\end{eqnarray*}
and the MPO at row~$r$ is
\begin{displaymath}
\begin{split}
U_r = \sum_{\begin{subarray}{1} \tilde{u}_1,...,\tilde{u}_L=1\\ \tilde{d}_1,...,\tilde{d}_L=1 \end{subarray}}^{D^2} & \tr \Big( \big[E_{r1}\big]^{\tilde{u}_1 \tilde{d}_1} \cdots \big[E_{rL}\big]^{\tilde{u}_L \tilde{d}_L} \Big) \times\\
& \times \ket{\tilde{u}_1,...,\tilde{u}_L} \bra{\tilde{d}_1,...,\tilde{d}_L}.
\end{split}
\end{displaymath}
In terms of these MPS and MPO, the scalar--product is a product of
MPO and MPS:
\begin{displaymath}
\scal{\Psi_A}{\Psi_A} = \bra{U_L} U_{L-1} \cdots U_2 \ket{U_1}
\end{displaymath}
The evaluation of this expression is, of course, intractable. With
each multiplication of a MPO with a MPS, the virtual dimension
increases by a factor of~$D^2$. Thus, after~$L$ multiplications,
the virtual dimension is~$D^{2L}$ -- which is exponential in the
number of rows. The expression, however, reminds of the
time--evolution of a
MPS~\cite{verstraeteripoll04,daley04,vidal04}. There, each
multiplication with a MPO corresponds to one evolution--step. The
problem of the exponential increase of the virtual dimension is
circumvented by restricting the evolution to the subspace of MPS
with a certain virtual dimension~$\tilde{D}$. This means that
after each evolution--step the resulting MPS is approximated by
the "nearest" MPS with virtual dimension~$\tilde{D}$. This
approximation can be done efficiently, as shown
in~\cite{verstraeteripoll04}. In this way, also
$\scal{\Psi_A}{\Psi_A}$ can be calculated efficiently: first, the
MPS $\ket{U_2}$ is formed by multiplying the MPS $\ket{U_1}$ with
MPO $U_2$. The MPS $\ket{U_2}$ is then approximated by
$\ket{\tilde{U}_2}$ with virtual dimension~$\tilde{D}$. In this
fashion the procedure is continued until $\ket{\tilde{U}_{L-1}}$
is obtained. The scalar--product $\scal{\Psi_A}{\Psi_A}$ is then
simply
\begin{displaymath}
\scal{\Psi_A}{\Psi_A} = \scal{U_L}{\tilde{U}_{L-1}}.
\end{displaymath}
The calculation of the coefficient matrix $\mathcal{N}_i$ is
closely related to the calculation of $\scal{\Psi_A}{\Psi_A}$:
$\mathcal{N}_i$ relies on the contraction of all but one of the
tensors $E_j$ according to the same scheme as before. The one
tensor that has to be omitted is $E_i$ -- the tensor related to
site~$i$. Assuming this contraction is performed by the
function~$\mathcal{G}_i(\cdot)$, $\mathcal{N}_i$ can be written as
\begin{displaymath}
\big[ \mathcal{N}_i \big]_{lrud}^{k}
{\phantom{\big]}}_{k'}^{l'r'u'd'} = \mathcal{G}_i \big(
E_1,...,E_M \big)_{lrud}^{l'r'u'd'} \delta_{k'}^k.
\end{displaymath}
If we join the indices $(klrud)$ and $(k'l'r'u'd')$, we obtain the
$d D^4 \times d D^4$--matrix that fulfills
equation~(\ref{eqn:scalaa}). To evaluate $\mathcal{G}_i(\cdot)$
efficiently, we proceed in the same way as before by interpreting
the rows in the contraction--structure as MPS and MPO. First, we
join all rows that lie above site~$i$ by multiplying the topmost
MPS~$\ket{U_1}$ with subjacent MPO and reducing the dimension
after each multiplication to~$\tilde{D}$. Then, we join all rows
lying below~$i$ by multiplying~$\bra{U_L}$ with adjacent MPO and
reducing the dimension as well. We end up with two MPS of virtual
dimension~$\tilde{D}$ -- which we can contract efficiently with
all but one of the tensors~$E_j$ lying in the row of site~$i$.

The scalar--product $\scal{\Psi_A}{\Psi_B}$ and the inhomogenity
$\mathcal{W}_i$ are calculated in an efficient way following the
same ideas. First, the $D D_B \times D D_B \times D D_B \times D
D_B$--tensors
\begin{displaymath}
\big[ F_j \big]_{(ll')(rr')}^{(uu')(dd')} = \sum_{k=1}^d \big[
A_j^* \big]_{lrud}^k \big[ B_j \big]_{l'r'u'd'}^k.
\end{displaymath}
are defined. The scalar--product $\scal{\Psi_A}{\Psi_B}$ is then
obtained by contracting all tensors~$F_j$ according to the
previous scheme -- which is performed by the
function~$\mathcal{F}(\cdot)$:
\begin{displaymath}
\scal{\Psi_A}{\Psi_B} = \mathcal{F} \big( F_1,...,F_M \big)
\end{displaymath}
The inhomogenity $\mathcal{W}_i$ relies on the contraction of all
but one of the tensors $F_j$, namely the function $\mathcal{G}_i
\big(\cdot)$, in the sense that
\begin{displaymath}
\big[ \mathcal{W}_i \big]_{lrud}^k = \sum_{l'r'u'd'=1}^D
\mathcal{G}_i \big( F_1,...,F_M \big)_{lrud}^{l'r'u'd'}
\big[B_i\big]_{l'r'u'd'}^k.
\end{displaymath}
Joining all indices $(klrud)$ in the resulting tensor leads to the
vector of length $d D^4$ that fulfills
equation~(\ref{eqn:scalab}). Thus, both the scalar--product
$\scal{\Psi_A}{\Psi_B}$ and the inhomogenity $\mathcal{W}_i$ are
directly related to the expressions $\mathcal{F}\big( F_1,...,F_M
\big)$ and $\mathcal{G}_i \big( F_1,...,F_M \big)$. These
expressions, however, can be evaluated efficiently using the
approximate method from before.

Summing up, we have an algorithm that allows the efficient
reduction of the virtual dimension of a PEPS - and thus the
efficient simulation of a time--evolution step within the subset of
PEPS.


\section{The Model: hard--core bosons in a 2--D~optical lattice} \label{sec:model}

This algorithm we use to study a system of bosons in a 2--D
optical lattice of size~$L \times L$. This system is characterized
by the Bose--Hubbard Hamiltonian
\begin{displaymath}
H = -J \sum_{<i,j>} \big( \adj{a_i} a_j +h.c.\big) + \frac{U}{2} \sum_i \hat{n}_i (\hat{n}_i -1)
+ \sum_i V_i \hat{n}_i,
\end{displaymath}
where $\adj{a}_i$ and $a_i$ are the creation and annihilation
operators on site~$i$ and $\hat{n}_i=\adj{a}_i a_i$ is the number
operator. This Hamiltonian describes the interplay between the
kinetic energy due to the next-neighbor hopping with amplitude~$J$
and the repulsive on-site interaction~$U$ of the particles. The
last term in the Hamiltonian models the harmonic confinement of
magnitude $V_i = V_0 (i-i_0)^2$. Since the total number of
particles $\hat{N}=\sum_i \hat{n}_i$ is a symmetry of the
Hamiltonian, the ground--state will have a fixed number of
particles~$N$. We choose this number by appending the term $-\mu
\hat{N}$ to the Hamiltonian and tuning the chemical
potential~$\mu$. The variation of the ratio~$U/J$ drives a
phase-transition between the Mott-insulating and the superfluid
phase, characterized by localized and delocalized particles
respectively~\cite{fisher89}. Experimentally, the variation
of~$U/J$ can be realized by tuning the depth of the optical
lattice~\cite{jaksch98,buechler03}. The quantity that is typically
measured is the momentum distribution. The is done by letting the
atomic gas expand and measuring the density distribution of the
expanded cloud. Thus, we will be mainly interested here in the
(quasi)--momentum distribution
\begin{displaymath}
n_k = \frac{1}{L^2} \sum_{r,s} \expect{\adj{a}_r a_s} e^{i 2 \pi k \cdot (r-s)/L^2}
\end{displaymath}
of the particles.

In the following, we focus on the limit of a hard--core
interaction, $U/J \to \infty$. In this limit, two particles are
prevented from occupying a single site. This limit is especially
interesting in one dimension where the particles form the
so--called Tonks-Girardeau gas~\cite{girardeau60,paredes04}. The
particles in this gas are strongly correlated -- which leads to
algebraically decaying correlation functions. In two dimensions,
the model was studied in detail in~\cite{aizenman04}. In the
hard--core limit, the Bose--Hubbard model is equivalent to a
spin--system with $XX$--interactions described by the Hamiltonian
\begin{displaymath}
H = -\frac{J}{2} \sum_{<i,j>} \big( \sx{i} \sx{j} + \sy{i} \sy{j} \big) +
\frac{1}{2} \sum_i \big( V_i-\mu \big) \sz{i}.
\end{displaymath}
Here, $\sx{i}$, $\sy{i}$ and $\sz{i}$ denote the Pauli-operators
acting on site~$i$. This Hamiltonian has the structure we can
simulate with the algorithm: it describes $L^2$ physical systems
of dimension~$d=2$ on a $L \times L$--square lattice.


\section{Application of the algorithm to hard--core bosons in a 2--D~lattice} \label{sec:application}

The principle of simulating a time--evolution step according to
$XX$--Hamiltonian is as follows: first, a PEPS $\ket{\Psi_A^0}$
with physical dimension~$d=2$ and virtual dimension~$D$ is chosen
as a starting state. This state is evolved by the time--evolution
operator~$U=e^{-i H \delta t}$ (we assume $\hbar=1$) to yield
another PEPS $\ket{\Psi_B}$ with increased virtual
dimension~$D_B$:
\begin{displaymath}
\ket{\Psi_B} = U \ket{\Psi_A^0}
\end{displaymath}
The virtual dimension of this state is then reduced to~$D$ by
applying the algorithm of the previous section. This means, a new
PEPS $\ket{\Psi_A}$ with virtual dimension~$D$ is calculated that
has minimal distance to $\ket{\Psi_B}$. This new PEPS is then the
starting state for the next time--evolution step.

The operator~$U$, however, increases the virtual dimension of a
PEPS by a factor that scales exponentially with~$L$. This is why
it is more convenient to approximate~$U$ by an operator that
increases the virtual dimension merely by a constant
factor~$\eta$. This is done by means of a Trotter--approximation:
first, the interaction--terms are classified in \emph{horizontal}
and \emph{vertical} according to their orientation and in
\emph{even} and \emph{odd} depending on whether the interaction is
between even--odd or odd--even rows (or columns). The Hamiltonian
can then be decomposed into a \emph{horizontal--even}, a
\emph{horizontal--odd}, a \emph{vertical--even} and a
\emph{vertical--odd} part:
\begin{displaymath}
H = H_{he} + H_{ho} + H_{ve} + H_{vo}
\end{displaymath}
The single--particle operators of the Hamiltonian can simply be
incorporated in one of the four parts. Using the
Trotter--approximation, the time--evolution operator~$U$ can be
written as a product of four evolution--operators:
\begin{equation} \label{eqn:uapprox}
U = e^{-i H \delta t} \approx
e^{-i H_{he} \delta t} e^{-i H_{ho} \delta t} e^{-i H_{ve} \delta t} e^{-i H_{vo} \delta t}
\end{equation}
Since each of the four parts of the Hamiltonian consists of a sum
of commuting terms, each evolution--operator equals a product of
two--particle operators
\begin{displaymath}
w_{ij} = e^{i \frac{\delta t J}{2} \big( \sx{i} \sx{j} + \sy{i} \sy{j} \big) }
\end{displaymath}
acting on neighboring sites~$i$ and~$j$. These two--particle
operators have a Schmidt--decomposition consisting of~$4$ terms:
\begin{displaymath}
w_{ij} = \sum_{\rho=1}^{4} u_i^{\rho} \otimes v_j^{\rho}
\end{displaymath}
One such two--particle operator~$w_{ij}$ applied to the
PEPS~$\ket{\Psi_A^0}$ modifies the tensors~$A_i^0$ and~$A_j^0$
associated with sites~$i$ and~$j$ as follows: assuming the
sites~$i$ and~$j$ are horizontal neighbors, $A_i^0$ has to be
replaced by
\begin{displaymath}
\big[B_i\big]^k_{l(r\rho)ud} = \sum_{k'=1}^2 \big[u_{i}^{\rho}\big]^k_{k'} \big[A_i^0\big]^{k'}_{lrud}
\end{displaymath}
and $A_j^0$ becomes
\begin{displaymath}
\big[B_j\big]^k_{(l\rho)rud} = \sum_{k'=1}^2 \big[v_{j}^{\rho}\big]^k_{k'} \big[A_j^0\big]^{k'}_{lrud}.
\end{displaymath}
These new tensors have a joint index related to the bond between
sites~$i$ and~$j$. This joint index is composed of the original
index of dimension~$D$ and the index~$\rho$ of dimension~$4$ that
enumerates the terms in the Schmidt--decomposition. Thus, the
effect of the two--particle operator~$w_{ij}$ is to increase the
virtual dimension of the bond between sites~$i$ and~$j$ by a
factor of~$4$. Consequently, $e^{-i H_{he} \delta t}$ and $e^{-i
H_{ho} \delta t}$ increase the dimension of every second
horizontal bond by a factor of~$4$; $e^{-i H_{ve} \delta t}$ and
$e^{-i H_{vo} \delta t}$ do the same for every second vertical
bond. By applying all four evolution--operators consecutively, we
have found an approximate form of the time--evolution operator~$U$
that -- when applied to a PEPS -- yields another PEPS with a
virtual dimension multiplied by a constant factor~$\eta=4$.

\begin{figure}[t]
    \begin{center}
        \includegraphics[width=0.44\textwidth]{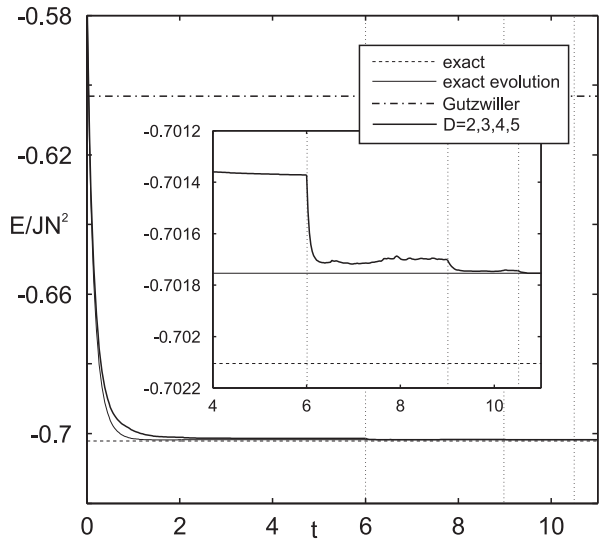}
     \end{center}
    \caption{
    Energy as a function of time for the imaginary time--evolution of
    the system of hard--core bosons on a $4 \times 4$--lattice.
    The evolutions are performed sequentially with
    PEPS of virtual dimension $D=2$, $D=3$, $D=4$ and $D=5$.
    The times at which $D$ is increased are indicated by
    vertical lines.
    For comparison, the exact ground state--energy,
    the exact imaginary time--evolution
    and the energy of the optimal Gutzwiller ansatz
    are included.
       }
    \label{fig:gstonks4}
\end{figure}

Even though the principle of simulating a time--evolution step has
been recited now, the implementation in this form is numerically
expensive.
This is why we append some notes about how to make the simulation more efficient:\\
\emph{1.- Partitioning of the evolution:} The number of required
numerical operations decreases significantly as one
time--evolution step is partitioned into~$4$ substeps: first the
state~$\ket{\Psi_A^0}$ is evolved by $e^{-i H_{vo} \delta t}$ only
and the dimension of the increased bonds is reduced back to~$D$.
Next, evolutions according to $e^{-i H_{ve} \delta t}$, $e^{-i
H_{ho} \delta t}$ and $e^{-i H_{he} \delta t}$ follow. Even though
the partitioning increases the number of evolution--steps by a
factor of~$4$, the number of multiplications in one evolution--step
decreases by a factor of~$\eta^3=64$.\\
\emph{2.- Optimization of the contraction order:} Most critical
for the efficiency of the numerical simulation is the order in
which the contractions are performed. We have optimized the order
in such a way that the scaling of the number of multiplications
with the virtual dimension~$D$ is minimal. For this, we assume
that the dimension~$\tilde{D}$ that tunes the accuracy of the
approximate calculation of $\mathcal{N}_i$ and $\mathcal{W}_i$ is
proportional to $D^2$, i.e. $\tilde{D}=\kappa D^2$. The number of
required multiplications is then of order~$\kappa^2 D^{12} L^2$
and the required memory scales as~$d \eta \kappa^2 D^8$.\\
\emph{3.- Optimization of the starting state:} The number of
sweeps required to reach convergence depends on the choice of the
starting state for the optimization. The idea for finding a good
starting state is to reduce the bonds with increased virtual
dimension~$\eta D$ by means of a Schmidt--decomposition. This is
done as follows: assuming the bond is between the horizontal
neighboring sites~$i$ and~$j$, the contraction of the tensors
associated with these sites, $B_i$ and $B_j$, along the bond
$i$--$j$ forms the tensor
\begin{displaymath}
\big[\mathcal{M}_{ij}\big]^k_{lud} {\phantom{\big]}}^{k'}_{r'u'd'} =
\sum_{\rho=1}^{D\eta} \big[B_i\big]^k_{l \rho ud} \big[B_j\big]^{k'}_{\rho r'u'd'}.
\end{displaymath}
By joining the indices $(k l u d)$ and $(k' r' u' d')$, this
tensor can be interpreted as a $d D^3 \times d D^3$--matrix. The
Schmidt--decomposition of this matrix is
\begin{displaymath}
\mathcal{M}_{ij} = \sum_{\rho=1}^{d D^3} c_{\rho} \mathcal{A}_i^{\rho} \otimes \mathcal{A}_j^{\rho}
\end{displaymath}
with the Schmidt--coefficients $c_{\rho}$ ($c_{\rho} \geq 0$) and
corresponding matrices $\mathcal{A}_i^{\rho}$ and
$\mathcal{A}_j^{\rho}$. We can relate these matrices to a new pair
of tensors~$A_i^0$ and~$A_j^0$ associated with sites~$i$ and~$j$:
\begin{eqnarray*}
\big[A_i^0\big]^k_{l \rho ud} & = & \sqrt{c_{\rho}} \big[\mathcal{A}_i^{\rho}\big]_{lud}^k\\
\big[A_j^0\big]^k_{\rho rud} & = & \sqrt{c_{\rho}} \big[\mathcal{A}_j^{\rho}\big]_{rud}^k
\end{eqnarray*}
The virtual dimension of these new tensors related to the bond
between sites~$i$ and~$j$ is equal to the number of terms in the
Schmidt--decomposition. Since these terms are weighted with the
Schmidt--coefficients~$c_{\rho}$, it is justified to keep only
the~$D$ terms with coefficients of largest magnitude. Then, the
contraction of the tensors~$A_i^0$ and~$A_j^0$ along the bond
$i$--$j$ with dimension~$D$ yields a good approximation to the
true value $\mathcal{M}_{ij}$:
\begin{displaymath}
\big[\mathcal{M}_{ij}\big]^k_{lud} {\phantom{\big]}}^{k'}_{r'u'd'} \approx
\sum_{\rho=1}^{D} \big[A_i^0\big]^k_{l \rho ud} \big[A_j^0\big]^{k'}_{\rho r'u'd'}.
\end{displaymath}
This method applied to all bonds with increased dimension provides
us with the starting state for the optimization.


\begin{figure}[t]
    \begin{center}
        \includegraphics[width=0.44\textwidth]{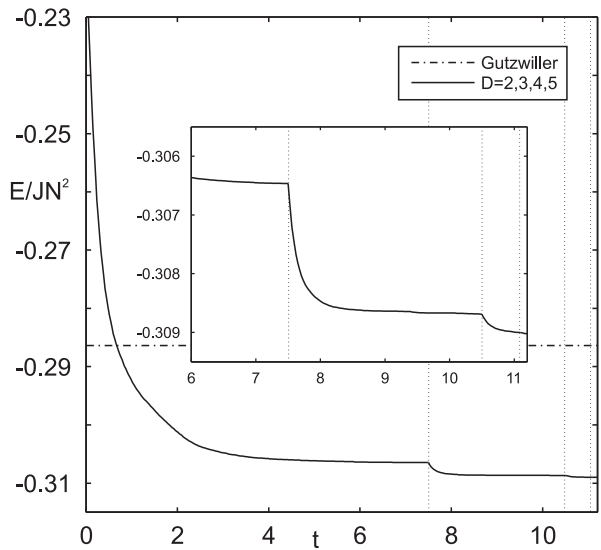}
     \end{center}
    \caption{
    Energy as a function of time for the imaginary time--evolution of
    the system of hard--core bosons on a $11 \times 11$--lattice.
    The evolutions are performed sequentially with
    PEPS of virtual dimension $D=2$, $D=3$, $D=4$ and $D=5$.
    The times at which $D$ is increased are indicated by
    vertical lines.
    For comparison, the energy of the optimal Gutzwiller ansatz is included.
       }
    \label{fig:gstonks11}
\end{figure}

\section{Ground state properties} \label{sec:groundstate}

In the following, we study the ground--state properties of the
system of hard--core bosons for lattice--sizes $4 \times 4$ and
$11 \times 11$. We calculate the ground--state by means of an
imaginary time--evolution which we can simulate with the method
from before.

We first focus on the $4 \times 4$--lattice for which we can
calculate the ground--state exactly and are able to estimate the
precision of the algorithm by comparison with exact results. In
fig.~\ref{fig:gstonks4}, the energy is plotted as the system
undergoes the imaginary time--evolution. We thereby assume a
time--step $\delta t = - i 0.03$. We choose the magnitude of the
harmonic confinement (in units of the tunneling--constant)
$V_0/J=36$. In addition, we tune the chemical potential
to~$\mu/J=3.4$ such that the ground state has
particle--number~$N=4$. With this configuration, we perform the
imaginary time--evolution both exactly and variationally with
PEPS. As a starting state we take a product state that represents
a Mott-like distribution with $4$~particles arranged in the center
of the trap and none elsewhere. The variational calculation is
performed with~$D=2$ first until convergence is reached; then,
evolutions with $D=3$, $D=4$ and $D=5$ follow. At the end, a state
is obtained that is very close to the state obtained by exact
evolution. The difference in energy is $| E_{D=5}-E_{exact} |
\backsimeq 6.4614 \cdot 10^{-5} J$. For comparison, also the exact
ground--state energy obtained by an eigenvalue--calculation and
the energy of the optimal Gutzwiller ansatz are included in
fig.~\ref{fig:gstonks4}. The difference between the exact result
and the results of the imaginary time--evolution is due to the
Trotter--error and is of order $O(\delta t^2)$. The energy of the
optimal Gutzwiller-Ansatz is well seperated from the exact
ground--state energy and the results of the imaginary
time--evolution.

In fig.~\ref{fig:gstonks11}, the energy as a function of time is
plotted for the imaginary time--evolution on the $11 \times
11$--lattice. Again, a time--step $\delta t = - i 0.03$ is assumed
for the evolution. The other parameters are set as follows: the
ratio between harmonic confinement and the tunneling constant is
chosen as $V_0/J=100$ and the chemical potential is tuned to
$\mu/J=3.8$ such that the total number of particles~$N$ is~$14$.
The starting state for the imaginary time--evolution is, similar
to before, a Mott-like distribution with $14$ particles arranged
in the center of the trap. This state is evolved within the
subset of PEPS with $D=2$, $D=3$, $D=4$ and $D=5$. As can be
gathered from the plot, this evolution shows a definite
convergence. In addition, the energy of the final PEPS lies well
below the energy of the optimal Gutzwiller ansatz.

\begin{figure}[t]
    \begin{center}
        \includegraphics[width=0.44\textwidth]{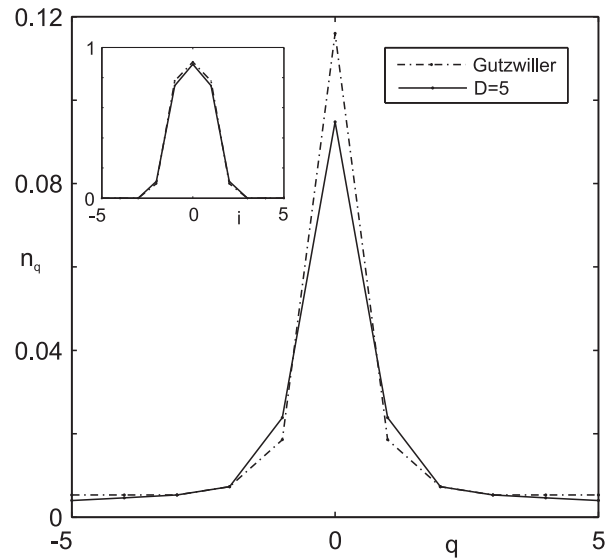}
     \end{center}
    \caption{
    (Quasi)--momentum distribution of the particles in the ground state
    of a $11 \times 11$--lattice. Plotted are results of the variational
    calculations with PEPS of dimension~$D=5$ and with the Gutzwiller ansatz.
    From the inset, the density of the particles can be gathered.
       }
    \label{fig:gsqmdist}
\end{figure}

The difference between the PEPS and the Gutzwiller ansatz becomes
more evident as one studies the momentum distribution of the
particles. The diagonal slice of the (quasi)--momentum
distribution is shown in fig.~\ref{fig:gsqmdist}. As can be seen,
there is a clear difference between the momentum distribution
derived from the PEPS and the one from the Gutzwiller ansatz. In
contrast, the PEPS and the Gutzwiller ansatz produce a very
similar density profile (see inset). The acceptability of the
Gutzwiller ansatz is due to the inhomegenity of the system: the
different average particle number at each site is the cause for
the correlations between different sites. These correlations are,
in many cases, good approximations. In contrast, the average
particle number is constant in homogeneous systems -- which leads
to correlations that are constant. Thus, the Gutzwiller ansatz is
expected to be less appropriate for the study of correlations of
homogeneous systems.


\section{Dynamics of the system} \label{sec:dynamics}

We now focus on the study of dynamic properties of hard--core
bosons on a lattice of size $11 \times 11$. We investigate the
responses of this system to sudden changes in the parameters and
compare our numerical results to the results obtained by the
Gutzwiller ansatz. The property we are interested in is the
fraction of particles that are condensed. For interacting and
finite systems, this property is measured best by the condensate
density~$\rho$ which is defined as largest eigenvalue of the
correlation--matrix~$\expect{\adj{a_i} a_j}$.


\begin{figure}[t]
    \begin{center}
         \includegraphics[width=0.44\textwidth]{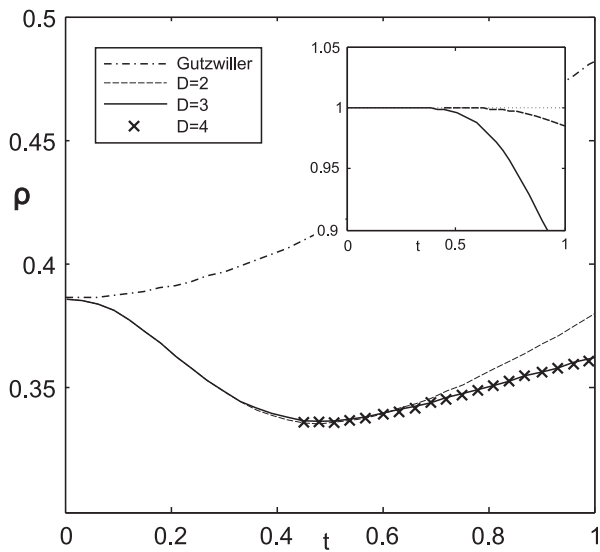}
     \end{center}
    \caption{
    Time evolution of the condensate density
    after a sudden change of the magnitude of the trapping potential
    from $V_0/J = 100$ to $V_0/J = 64$.
    As a starting state, we use the Gutzwiller--approximation of the ground state.
    The evolution is performed
    on the basis of the Gutzwiller ansatz and PEPS with $D=2$, $D=3$ and $D=4$.
    From the inset, the overlap between the PEPS with $D=2$ and
    $D=3$ (solid line) and the PEPS with $D=3$ and $D=4$ (dashed
    line) can be gathered.
       }
    \label{fig:evolution}
\end{figure}

First, we study the time evolution of the condensate density after
a sudden change of the trapping potential. We start with a
Gutzwiller--approximation of the ground state in case of a
trapping potential of magnitude $V_0/J = 100$. The chemical
potential we tune to $\mu/J=3.8$ to achieve an average
particle--number of $\expect{\hat{N}}=14$. This state we expose to
a trapping potential of magnitude $V_0/J = 64$ and calculate the
evolution of the condensate density using the Gutzwiller ansatz
and PEPS with $D=2$, $D=3$ and $D=4$. We thereby assume a
time--step $\delta t = 0.03$. To assure that our results are
accurate, we proceed as follows: first, we perform the simulation
using PEPS with $D=2$ and $D=3$ until the overlap between these
two states falls below a certain value. Then, we continue the
simulation using PEPS with $D=3$ and $D=4$ as long as the overlap
between these two states is close to~$1$. The results of this
calculation can be gathered from fig.~\ref{fig:evolution}. What
can be observed is that the results obtained from using PEPS are
qualitatively very different from the result based on the
Gutzwiller ansatz. The inset in fig.~\ref{fig:evolution} shows the
overlap of the $D=2$ with the $D=3$--PEPS and the $D=3$ with the
$D=4$--PEPS.


\begin{figure}[t]
    \begin{center}
         \includegraphics[width=0.44\textwidth]{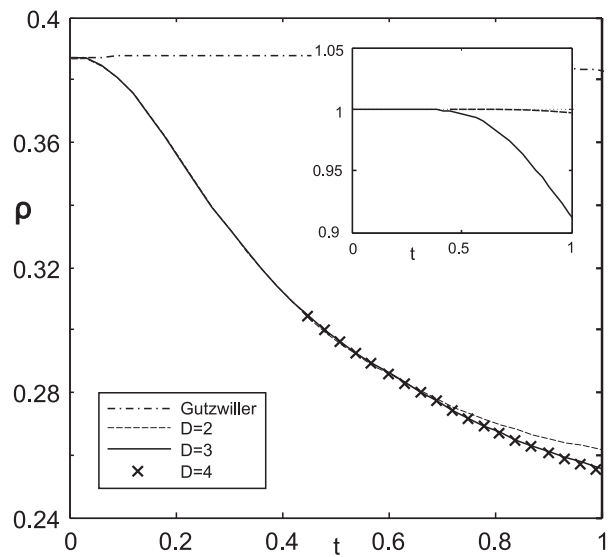}
     \end{center}
    \caption{
    Time evolution of the condensate density
    after a sudden shift of the center of the trap
    by one site in $x$-- and $y$--direction.
    Starting state is the Gutzwiller--approximation of the ground state.
    The evolution is performed using the Gutzwiller ansatz
    and PEPS with $D=2$, $D=3$ and $D=4$.
    The inset shows the overlap between the PEPS with
    $D=2$ and $D=3$ (solid line) and $D=3$ and $D=4$ (dashed
    line).
       }
    \label{fig:destruction}
\end{figure}

In fig.~\ref{fig:destruction}, the time--evolution of the
condensate density after a sudden shift of the trapping potential
is plotted. As a starting state, again the
Gutzwiller--approximation of the ground state in a trap of
magnitude $V_0/J = 100$ is used. This state is evolved with
respect to a trapping potential that is shifted by one
lattice--site in $x$-- and $y$--direction. We assume a time--step
$\delta t = 0.03$ and tune the chemical potential to $\mu/J =
3.8$. As before, we perform the simulation successively with
$D=2$, $D=3$ and $D=4$ and judge the accuracy of the results by
monitoring the overlap between PEPS with different~$D$s. From the
plot, it can be gathered that the evolution of the condensate
density based on the Gutzwiller ansatz is qualitatively again very
different from the evolution obtained from using PEPS. The
evolution obtained from using PEPS shows a definite damping. The
shift of the trap thus provokes a destruction of the condensate.
The evolution based on the Gutzwiller ansatz doesn't show this
feature.


\begin{figure}[t]
    \begin{center}
         \includegraphics[width=0.44\textwidth]{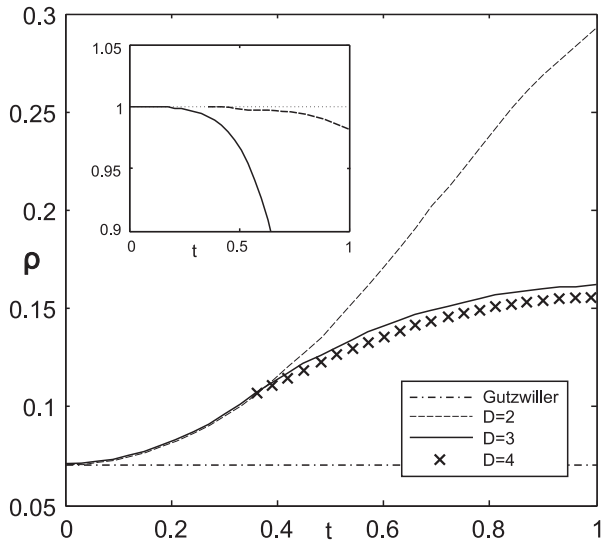}
     \end{center}
    \caption{
    Time evolution of the condensate density
    starting from a Mott--distribution
    with $14$--particles arranged in the center of the trap.
    The magnitude of the trapping potential is $V_0/J = 100$.
    For the evolution, the Gutzwiller ansatz and PEPS with $D=2$, $D=3$ and $D=4$ are used.
    The inset shows the overlap between the $D=2$ and $D=3$--PEPS
    (solid line) and the $D=3$ and $D=4$--PEPS (dashed line).
    }
    \label{fig:creation}
\end{figure}

As a contrary example, we study the evolution of a
Mott-distribution with $14$ particles arranged in the center of
the trap. We assume $V_0/J=100$, $\mu/J=3.8$ and $\delta t =
0.03$. We perform the simulation in the same way as before with
$D=2$, $D=3$ and $D=4$. In fig.~\ref{fig:creation}, the time
evolution of the condensate density is plotted. It can be observed
that there is a definite increase in the condensate fraction. The
Gutzwiller ansatz is in contrast to this result since it predicts
that the condensate density remains constant.


\section{Accuracy and performance of the algorithm} \label{sec:performance}

\begin{figure}[t]
    \begin{center}
         \includegraphics[width=0.44\textwidth]{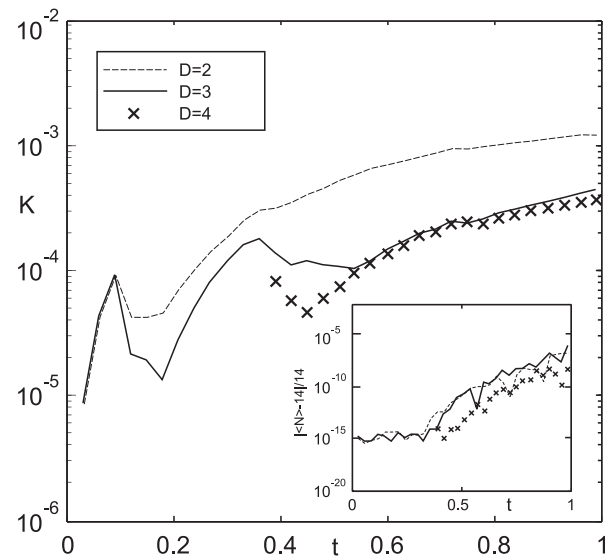}
     \end{center}
    \caption{
    Distance $K$ between the time--evolved state and the state with
    reduced virtual dimension. The virtual dimensions $D=2$, $D=3$ and $D=4$ are included.
    The distance is plotted for the
    evolution of a Mott-distribution with $N=14$, as explained in fig.~\ref{fig:creation}.
    From the inset, the deviation of the particle number from the value~$14$
    can be gathered.
    }
    \label{fig:kvalues}
\end{figure}

Finally, we make a few comments about the accuracy and the
performance of the algorithm. One indicator for the accuracy of
the algorithm is the distance between the time--evolved state and
the state with reduced virtual dimension. For the time--evolution
of the Mott--distribution that was discussed in
section~\ref{sec:dynamics}, this quantity is plotted in
fig.~\ref{fig:kvalues}. We find that the distance is typically of
order $10^{-3}$ for $D=2$ and of order $10^{-4}$ for $D=3$ and
$D=4$. Another quantity we monitor is the total number of
particles $\expect{\hat{N}}$. Since this quantity is supposed to
be conserved during the whole evolution, its fluctiations indicate
the reliability of the algorithm. From the inset in
fig.~\ref{fig:kvalues}, the fluctuations of the particle number in
case of the time--evolution of the Mott--distribution can be
gathered. We find that these fluctuations are at most of
order~$10^{-5}$.

The main bottleneck for the performance of the algorithm is the
scaling of the number of required multiplications with the virtual
dimension~$D$. As mentioned in section~\ref{sec:application}, the
number of required multiplications is of order~$D^{12}$. Our
simulations were run on a workstation with a $3.0$~GHz Intel Xeon
processor. On such a system, one evolution step on a $11 \times
11$--lattice with $D=5$ required a computing time of $55$~hours.
Another bottleneck for the algorithm forms the scaling of the
required memory with the virtual dimension~$D$ -- which is of
order $D^8$. The simulation on a $11 \times 11$--lattice with
$D=5$ thereby required a main memory of $2$~GB. These bottlenecks
make it difficult at the moment to go beyond a virtual dimension
of~$D=5$. Nonetheless, a virtual dimension of~$D=5$ is expected to
yield good results for many problems already. We intend to
overcome the limitations of time and space by distributing tensor
contractions among several processors in a future project.


\section{Conclusions}

Summing up, we have studied the system of hard--core bosons on a
2--D lattice using a variational method based on PEPS. We have
thereby investigated the ground state properties of the system and
its responses to sudden changes in the parameters. We have
compared our results to results based on the Gutzwiller ansatz. We
have observed that the Gutzwiller ansatz predicts very well the
density distribution of the particles. However, the momentum
distribution obtained from the Gutzwiller ansatz is, though
qualitatively similar, quantitatively clearly different from the
distribution obtained from the PEPS ansatz. In addition, the PEPS
and the Gutzwiller ansatz are very different in the prediction of
time evolutions. We conclude that the Gutzwiller ansatz has to be
applied carefully in these cases. The simulations done in this
paper give a clear demonstration of the power of the
PEPS-approach, both for finding ground states in
higher-dimensional quantum spin systems and for simulating
real-time evolution.


\end{document}